\documentclass[traditabstract]{aa}

\usepackage{array}
\usepackage{amsmath}
\usepackage{natbib}
\usepackage{graphicx}
\usepackage{xspace}
\usepackage{float}
\usepackage{xcolor}
\usepackage[hyphens]{url}

\bibpunct{(}{)}{;}{a}{}{,}

\newcommand{\microns}{\,$\mu$m\xspace}
\newcommand{\mm}{\,mm\xspace}
\newcommand{\yr}{\,yr\xspace}
\newcommand{\mdote}{\dot{M}_{\rm env}}

\newcommand{\mstar}{M_\star}
\newcommand{\mdisk}{M_{\rm disk}}

\newcommand{\tstar}{T_\star}
\newcommand{\rsun}{R_\odot}

\newcommand{\msun}{M_\odot}
\newcommand{\rstar}{R_\star}
\newcommand{\rmind}{R^{\rm disk}_{\rm min}}
\newcommand{\rmaxd}{R^{\rm disk}_{\rm max}}
\newcommand{\rmine}{R^{\rm env}_{\rm min}}

\newcommand{\rc}{R_{\rm c}}

\newcommand{\rsub}{R_{\rm sub}}

\begin{document}

\title{A modular set of synthetic spectral energy distributions for young stellar objects}

\author{Thomas P. Robitaille\inst{1,2,}\thanks{Visiting Researcher, School of Physics \& Astronomy, The University of Leeds, Leeds LS2 9JT, United Kingdom
}}
\institute{Max Planck Institute for Astronomy, K\"onigstuhl 17, Heidelberg 69117, Germany
\and Headingley Enterprise and Arts Centre, Bennett Road, Leeds, LS6 3HN, United Kingdom\\
\email{thomas.robitaille@gmail.com} 
}

\date{Received 8 December 2014 / Accepted 3 August 2016}

\abstract{In this paper, I present a new set of synthetic spectral energy distributions (SEDs) for young stellar objects (YSOs) spanning a wide range of evolutionary stages, from the youngest deeply embedded protostars to pre-main-sequence stars with few or no disks. These models include significant improvements on the previous generation of published models: in particular, the new models cover a much wider and more uniform region of parameter space, do not include highly model-dependent parameters, and include a number of improvements that make them more suited to modeling far-infrared and sub-mm observations of forming stars. Rather than all being part of a single monolithic set of models, the new models are split up into sets of varying complexity. The aim of the new set of models is not to provide the most physically realistic models for young stars, but rather to provide deliberately simplified models for initial modeling, which allows a wide range of parameter space to be explored. I present the design of the model set, and show examples of fitting these models to real observations to show how the new grid design can help us better understand what can be determined from limited unresolved observations. The models, as well as a Python-based fitting tool are publicly available to the community.}{}


\titlerunning{A modular set of synthetic SEDs for YSOs}
\authorrunning{Robitaille}

\maketitle

\section{Introduction}

Over the last two decades, increasingly sensitive and detailed surveys of our
Galaxy at infrared wavelengths have resulted in a dramatic increase in the
number of known young stellar objects (YSOs). The \textit{Spitzer}/GLIMPSE
survey at mid-infrared wavelengths contains tens of thousands of YSOs, with
the sample of around 10,000 YSO candidates from \citet{robitaille:08:2413}
representing only the brightest ones. \textit{Spitzer} has also revealed thousands of YSOs in a number of individual regions, including over 8,000 YSO candidates in the survey of Cygnus
X North \citep{beerer:10:679}, over 1,000 YSO candidates in the c2d survey of nearby
star-forming regions \citep{evans:09:321}, and almost 3,500 YSO in the Orion A and B molecular clouds \citep{megeath:12:192}. The WISE survey \citep{wright:10:1868}, while less sensitive than \textit{Spitzer}, covers the whole sky and has already unveiled thousands of YSOs \citep{koenig:14:131}. \textit{Herschel} has both provided long-wavelength data for known YSO candidates,
and has also revealed new extremely embedded protostars \citep[e.g.,][]{stutz:13:36}.

While the Hubble Space Telescope (HST) and the Atacama Large Millimeter Array (ALMA)
allow the study of a smaller sample of YSOs with exquisite resolution and
sensitivity, the vast majority of YSOs in the Galaxy will remain unresolved for
a while yet. Nevertheless, with such a large sample of known YSOs and YSO candidates, we can
make significant progress in quantifying the formation of stars across a wide range of environments. Extracting physical properties from limited --
often unresolved -- observations however requires radiative transfer modeling.

In \citet[][hereafter R06]{robitaille:06:256}, we presented a set of approximately 20,000
radiative transfer models, each of which had spectral energy distributions (SEDs) computed for ten viewing angles
and 50 apertures. The models were set up by first sampling the central source
mass between 0.1 and 50$\msun$, and the central source age between $10^3$ and
$10^7$\yr (excluding post-main-sequence objects), and using evolutionary tracks
to derive the source temperature and radius in each case. The parameters for
the circumstellar environment, consisting of an accretion disk, infalling
envelope, and bipolar cavities, were then sampled from ranges that were
functions of the stellar mass and/or age. In total, 14 parameters were used to
define the models.

In \citet[][hereafter R07]{robitaille:07:328}, we presented a tool to rapidly
fit model SEDs from R06 to observations, correctly taking into account the effects of
extinction and apertures. The aim of the tool was not to identify the best-fitting model, but rather the ensemble of models that could provide a good fit, thereby identifying likely ranges in parameters space.

The R06 models and the R07 fitting tool have been extensively used to model the
SEDs of thousands of sources across many regions of star formation. These models have been invaluable in a number of studies to learn about the physical properties
of young stars and regions. For example, \citet{forbrich:10:1453} carefully
modeled the SEDs of a number of protostars in the IC348 and NGC2264
star-forming regions, demonstrating that it was crucial to include the
intra-cluster extinction in order to reliably determine evolutionary stages of
young sources in embedded clusters. Another example is the study by \citet{mottram:11:A149}, which made use of the R06 models in order to determine bolometric fluxes and luminosities for over a thousand embedded massive stars in the Red MSX Source (RMS) survey.

In this paper, I present a new set of model SEDs for YSOs that significantly
improves on the R06 models. Despite the success of the R06 models, a number of
issues with those models remain, and these are described in Section
\ref{sec:limitations}. I then present an overview of the setup and methods for the new set of models in Sections~\ref{sec:overview} and \ref{sec:methods} respectively, and present some initial results derived from the new models in Section~\ref{sec:results}. I discuss remaining caveats relating to the models in Section~\ref{sec:caveats}, and limitations inherent to SED modeling in Section~\ref{sec:general_limitations}.

\section{Limitations of the previous models}

\label{sec:limitations}

In this section, I review the main limitations of the R06 models in order to  justify the choices made for the new set of models. For a
discussion of more general issues inherent in any SED modeling, the reader is referred to Section~\ref{sec:general_limitations}.

\subsection{Model-dependent parameters}

Of the 14 parameters defining the R06 models, some required specific assumptions
to be made in order to be used in the analytical description of the model. For
instance, the stellar mass and age had no direct impact on the SED,
but instead, evolutionary tracks were needed to transform these into the
stellar temperature and radius, which are the properties that actually have an
effect on the SED. When fitting these models to observed SEDs, the parameters
we can directly determine in a best-case scenario are the stellar temperature and luminosity
(and therefore radius). The stellar mass and age are then only determined with
the assumption of the evolutionary tracks. Similarly, the envelope infall rate
was converted to an envelope density structure via the assumption of the Ulrich
(1976) collapse model. Therefore, when one derives stellar masses, ages, and
envelope infall rates from the R06 models, one is implicitly assuming a
specific set of evolutionary tracks and a collapse model. However, users of the
models may not necessarily realize the inherent assumptions, and most
importantly cannot easily change these assumptions since they are built into the models.

\subsection{Unevenly sampled parameter space}

\label{sec:uneven_sampling}

The R06 models are defined by sampling circumstellar environment parameters
from ranges that are functions of the stellar mass and age. For example, the envelope infall rate is sampled from ranges that decrease over time, while the cavity opening angle is sampled from ranges that increase with time. The motivation for these was to restrict the parameter space coverage to regions that were thought to be realistic in order to minimize the computing time required, but the downside was that this led to correlations between parameters in the models before they were even applied
to observations.

Examples include a correlation between the envelope infall
rate $\mdote$ and stellar mass $\mstar$ (since the envelope infall rate is sampled from a
constant range in $\mdote/\mstar$), between the envelope infall rate and the disk
accretion rate (since both are sampled from ranges that decrease for larger
ages), and an anti-correlation between infall rate and cavity opening angle (since, as mentioned above, the latter is sampled from a range that increases with time). These correlations can be seen in Figure 1 of \cite{robitaille:08:290}.
These built-in correlations mean that it is very difficult to look for such
trends in samples of objects, since the built-in trends would mask
any real trend.

\subsection{Fixed model complexity}

The R06 models, defined by 14 parameters, are often used without testing
simpler models. For example, all of the models with envelopes in the R06 set also
have disks, and furthermore the higher the envelope density, the higher the
minimum disk mass found in the models. This means that there are, in fact, no
deeply embedded protostellar models that do not have a disk. This in turn
implies that it is not possible to use the models to try and find evidence for
disks in embedded YSOs because there are no models to test the hypothesis that there is no disk.

\subsection{Lack of cold dust at long wavelengths}

The outer radii of the envelopes in the R06 models were defined as the radius where the temperature would drop to 30K if the envelope was optically thin to radiation. While for more embedded YSOs, the temperature may drop below 30K at the outer radius, those models only guarantee that they include all dust hotter than 30K. This in turn means that the models may not include enough cold material compared to observed sources, and may therefore show a deficit of long-wavelength emission.

The reason for this limitation was that the models were primarily designed with the \textit{Spitzer} IRAC/MIPS and shorter wavelength data in mind, and at the time there were no surveys equivalent to the \textit{Herschel} data, which now provides high-resolution long-wavelength fluxes for thousands of YSOs.

As a result, investigations using \textit{Herschel} data have predictably found that the models are not always able to fit at the longest wavelengths. For example, \citet{Sewilo:10:L73} found that some of the \textit{Herschel} sources in the Large Magellanic Cloud were not well fit by the R06 models beyond 100\microns. Because the R06 models do not include much dust below 30K, none of the model SEDs peak at wavelengths longer than 100\microns, but the coldest protostars observed with \textit{Herschel} typically peak around 200 or 300\microns. While the R06 models were well-suited to \textit{Spitzer} observations, new models are required now that \textit{Herschel} observations of YSOs are common.

\subsection{Poor signal-to-noise at long wavelengths}

\label{sec:lowsn}

The R06 models were computed using the Monte-Carlo radiative transfer code developed by \citet{whitney:03:1079,whitney:03:1049,whitney:04:1177}, which at the time computed the long-wavelength SED by sampling equal energy photon packets in the same way as for shorter wavelengths. However, traditional Monte-Carlo sampling done in this way results in the number of photons emitted being highest where the SED is brightest, and conversely lower where the SED is fainter, such as in the far-infrared and at mm wavelengths. As a result, most SEDs in the R06 set had poor signal-to-noise (S/N) at 1\mm and beyond, and some of the less embedded models even had poor S/N beyond 100\microns. While in many cases this has not been too problematic, in the sense that observations also have poor S/N where for the faintest sources, we ideally need models with high S/N at all wavelengths to make the most efficient use of available observations.

\section{Model components and set-up}

\label{sec:overview}

In this section, I present an overview of the new set of models, which addresses the issues described in Section~\ref{sec:limitations}, and includes a number of further improvements.

\subsection{Design philosophy}

\label{sec:overall}

The new models were not computed in a single monolithic set, but rather as
several sets of models with increasing complexity. For instance, one of the sets consists of models with only a star with a surrounding disk, another set includes a disk and an envelope, but no bipolar cavities, and yet another set includes a disk, envelope, and bipolar cavities.

This modularity allows us to ask which model offers the best representation of the data,
before even looking at the actual values of the parameters (see an example in Section~\ref{sec:fitting_examples}). For example, a
source might be fit by a complex model with a central source, a disk, and an
envelope with bipolar cavities, but it might be equally well fit by a model
with only a disk, indicating that there is no strong evidence one way or another for the presence of an envelope. It is important to assess not only the goodness of fit
but also the simplicity or complexity of the model, since it is easy but not
always meaningful to fit any set of data with an arbitrarily complex model.
This design can also allow the available sets of models to grow over time.

The components used to generate the model sets are described in Section~\ref{sec:components}. For components containing free parameters, the free parameters were  uniformly randomly sampled between a minimum and maximum value. The ranges of values used for each parameter are given in Table~\ref{table:parameters}. By randomly sampling in uniform ranges, we can avoid correlations between parameters which were present in the R06 models. On the other hand, some of the combinations of parameters may be unphysical -- since the definition of what might be considered physical will change over time (for example with stellar evolutionary tracks) the models presented here span a broad parameter space, and it is left to the user of the models to decide which models to ignore as being unphysical.

In general, models can be divided into two main categories. The first are models that aim to be as realistic as possible -- for example in the case of radiative transfer, models with a realistic 3-d distribution of dust and with the best available dust model (with variable dust properties depending on location and temperature). The second category of models are models that are simpler but aim to provide insight into the effect of various physical processes, components, and so on. The collection of models presented in this paper is firmly in the second category: the aim is not to provide the most realistic model, but rather simple models that can be used to explore large regions of parameter space. This influences some of the decisions outlined in the following sections. I encourage users to use these models as a starting point for modeling observational data of young stars, but to follow this up with more detailed modeling if the observations cannot be reproduced with the simple models, or if spatially resolved data or spectra are available.

\subsection{Components}

\label{sec:components}

As mentioned in \S\ref{sec:overall} the new models consist
of a combination of components which I describe in the following sections.

\subsubsection{Central source}

\label{sec:source}

The central source, present in all models, was set to be a spherical source.
Unlike the R06 models, the central source was not assigned a mass, and the stellar properties were not derived from evolutionary tracks. Instead, the central source was defined directly using a stellar radius $\rstar$ and effective
temperature $\tstar$. While this does result in some of the models having
unphysical combinations of $\rstar$ and $\tstar$, it allows the models to be
independent of stellar evolutionary tracks. The idea is that users of the models can then -- if needed -- assume a specific set of tracks, decide which models are physical according to those tracks, and assign masses and ages to the stars.

The effective temperature of the source was used to select appropriate stellar photosphere models. For
temperatures including and above 4,000\,K, the photosphere models from \citet{castelli:04} were used, while for temperatures below 4,000\,K, models
 computed with the PHOENIX code \citep{brott:05} and intended for the GAIA mission\footnote{\url{http://www.hs.uni-hamburg.de/EN/For/ThA/phoenix/gaia_info.html}} were used instead. Since the central source is not defined in terms of mass, we cannot calculate a surface
gravity $\log{[g]}$. However, $\log{[g]}$ does not have a large impact on
stellar photospheres in the range 3,000\,K to 20,000\,K, while above and below
these temperatures, the largest difference depending on the choice of stellar photosphere model
is generally of the order of 10\% relative to the $\log{[g]}=4.0$ models.
Therefore, the $\log{[g]}=4.0$ models were always used, with the caveat that
these may be wrong by up to 10\% for low and high temperatures, which affects
mostly the near-infrared fluxes for non-embedded models.

\begin{table*}
\caption{Parameter ranges sampled}
\label{table:parameters}
\centering
\begin{tabular}{|l|cccc|}
\hline
Parameter & Symbol & Minimum & Maximum & Sampling \\
\hline
Stellar radius & $\rstar$ & 0.1\,$\rsun$ & 100\,$\rsun$ & Log \\
Stellar temperature & $\tstar$ & 2000\,K & 30000\,K & Log \\
Disk mass [dust] & $\mdisk$ & $10^{-8}$\,$\msun$ & $0.1$\,$\msun$ & Log \\
Disk inner radius & $\rmind$ & $\rsub$ & 1000\,$\rsub$ & Log \\
Disk outer radius & $\rmaxd$ & 50\,AU & 5000\,AU & Log \\
Disk flaring power & $\beta$ & 1 & 1.3 & Linear \\
Disk surface density power & $p$ & $-2$ & $0$ & Linear \\
Disk scaleheight & $h_{\rm 100AU}$ & 1\,AU & 20\,AU & Log \\
Envelope density [dust] & $\rho_0^{\rm env}$ & $10^{-24}$\,g/cm$^3$ & $10^{-16}$\,g/cm$^3$ & Log \\
Envelope density power & $\gamma$ & $-2$ & $-1$ & Linear \\
Envelope centrifugal radius & $\rc$ & 50\,AU & 5000\,AU & Log \\
Cavity density [dust] & $\rho_0^{\rm cav}$ & $10^{-23}$\,g/cm$^3$ & $10^{-20}$\,g/cm$^3$ & Log \\
Cavity opening angle & $\theta_0$ & 0$^\circ$ & 60$^\circ$ & Linear \\
Cavity power & $c$ & 1 & 2 & Linear \\
\hline
\end{tabular}
\end{table*}

\subsubsection{Disk}

\label{sec:disk}

In this set of models, only passive disks are included. For
embedded YSOs, the effect of not explicitly including accretion is likely to be
minimal because the heating from viscous dissipation in the disk is typically
not very important, and the biggest effect is the increase in luminosity from
the star. Since the star is embedded, the radiation from the star gets reprocessed
and therefore the shape of the stellar spectrum is not important.

When modeling disks at near-infrared and longer wavelengths, the passive disks
can still be used to model accreting disks for a similar reason: while we
expect a little extra heating in the disk from the viscous dissipation, most of
the accretion luminosity comes from the stellar surface, and again the increase
in luminosity of the central source is the most important effect, so a model of
a passive disk with a higher central luminosity star will likely be adequate in
most cases.

Of course, the lack of accretion does mean that UV and optical fluxes for non-embedded sources with
strong accretion cannot reliably be modeled, since none of the passive
disk models will produce the adequate excess UV and optical emission typically
observed toward accreting sources.

Disks in hydrostatic equilibrium are expected to be flared \citep{shakura:73:337}, and at earlier times, when the dust is well coupled to the gas, the dust in the disk follows the same structure. Over time, as the dust settles to the mid-plane, the effective flaring for the disk may change. The flaring and the scaleheight of the disk is therefore parametrized such that it covers the range of flaring from hydrostatic disks to flat disks.

The density distribution of the passive flared disks is given in cylindrical polar coordinates $(R,z,\phi)$ by
\begin{equation}
\rho(R,z,\phi) = \rho_0^{\rm disk}\,\left(\frac{R_0}{R}\right)^{\beta - p}\,\exp{\left[-\frac{1}{2}\left(\frac{z}{h(R)}\right)^2\right]}
,\end{equation}
where $\rho_0^{\rm disk}$ is defined by the disk dust mass $\mdisk$, $p$ is the surface density radial exponent, $\beta$ is the disk flaring power, and the disk scaleheight $h(R)$ is given by
\begin{equation}
h(R) = h_0\,\left(\frac{R}{R_0}\right)^\beta.
\end{equation}
The disk is truncated at the inner radius $\rmind$ and the outer radius $\rmaxd$.

The free parameters varied for all models with a disk are the disk dust mass $\mdisk$, outer radius $\rmaxd$, flaring exponent $\beta$, surface density exponent $p$, and scaleheight $h_0$. For some of the models, the disk inner radius $\rmind$ was also varied (as described in \S\ref{sec:sampling}).

\subsubsection{Envelope}

Two types of envelopes were included in the models -- the first are
spherically-symmetric power-law envelopes, and the second are rotationally
flattened envelopes. The reason for including both types of envelopes is because this will
allow users to investigate the constraints on the envelope structure from
observations: while a model with a more complex rotationally flattened envelope
may fit well, it is important to also check whether a model with a simpler
spherically symmetric envelope can also fit. The spherically symmetric models also have more flexibility as to what power-law to use for the radial dependence of the density.

The power-law envelope density structure is given by
\begin{equation}
\rho(r) = \rho_0^{\rm env}\,\left(\frac{r}{r_0}\right)^\gamma
,\end{equation}
where $\rho_0^{\rm env}$ is the density of the envelope at the radius $r_0$,
and serves as the scaling for the envelope density, and $\gamma$ is the radial
power of the density.

\begin{table*}
\caption{Model sets presented in this paper}
\label{table:grids}
\centering
\begin{tabular}{|l|c|cccccc|cc|}
\hline
Model set & Icon & Star & Disk & Envelope & Cavity & Ambient & Inner radius & Variables & Models \\ 
\hline
\texttt{s---s-i} & \parbox[c]{1.3cm}{\includegraphics[width=1.3cm]{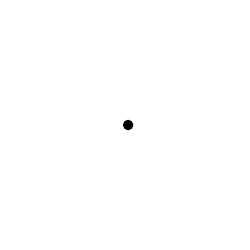}} & yes &    \ldots     &    \ldots     & \ldots  & \ldots & \ldots  & 2 & 10,000 \\
\texttt{sp--s-i} & \parbox[c]{1.3cm}{\includegraphics[width=1.3cm]{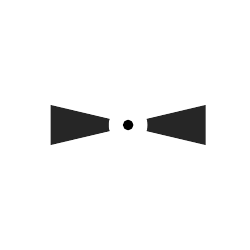}} & yes &  passive  &    \ldots     & \ldots  & \ldots  & $\rsub$  & 7 & 10,000 \\
\texttt{sp--h-i} & \parbox[c]{1.3cm}{\includegraphics[width=1.3cm]{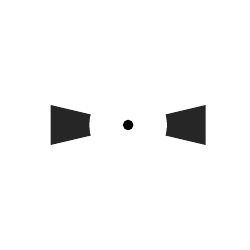}} & yes &  passive  &    \ldots     & \ldots  & \ldots  & variable & 8 & 10,000 \\
\texttt{s---smi} & \parbox[c]{1.3cm}{\includegraphics[width=1.3cm]{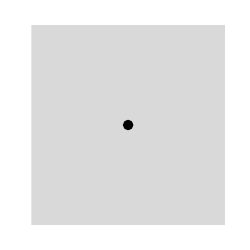}} & yes &    \ldots     &    \ldots     & \ldots  & yes & $\rsub$  & 2 & 10,000 \\
\texttt{sp--smi} & \parbox[c]{1.3cm}{\includegraphics[width=1.3cm]{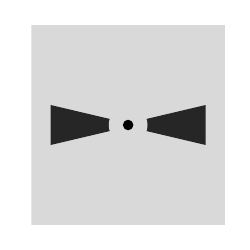}} & yes &  passive  &    \ldots     & \ldots  & yes & $\rsub$  & 7 & 10,000 \\
\texttt{sp--hmi} & \parbox[c]{1.3cm}{\includegraphics[width=1.3cm]{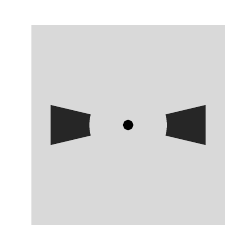}} & yes &  passive  &    \ldots     & \ldots  & yes & variable & 8 & 10,000 \\
\texttt{s-p-smi} & \parbox[c]{1.3cm}{\includegraphics[width=1.3cm]{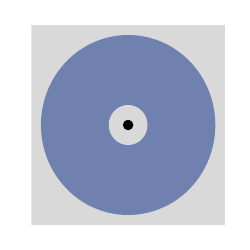}} & yes &    \ldots     & power-law & \ldots  & yes & $\rsub$  & 4 & 10,000 \\
\texttt{s-p-hmi} & \parbox[c]{1.3cm}{\includegraphics[width=1.3cm]{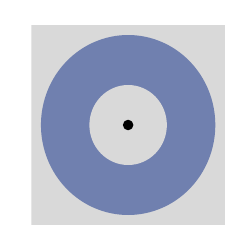}} & yes &    \ldots     & power-law & \ldots  & yes & variable & 5 & 10,000 \\
\texttt{s-pbsmi} & \parbox[c]{1.3cm}{\includegraphics[width=1.3cm]{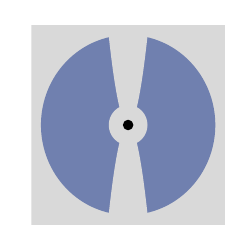}} & yes &    \ldots     & power-law & yes & yes & $\rsub$  & 7 & 10,000 \\
\texttt{s-pbhmi} & \parbox[c]{1.3cm}{\includegraphics[width=1.3cm]{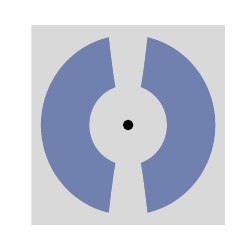}} & yes &    \ldots     & power-law & yes & yes & variable & 8 & 10,000 \\
\texttt{s-u-smi} & \parbox[c]{1.3cm}{\includegraphics[width=1.3cm]{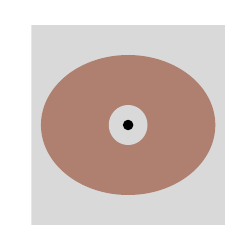}} & yes &    \ldots     &  Ulrich   & \ldots  & yes & $\rsub$  & 4 & 10,000 \\
\texttt{s-u-hmi} & \parbox[c]{1.3cm}{\includegraphics[width=1.3cm]{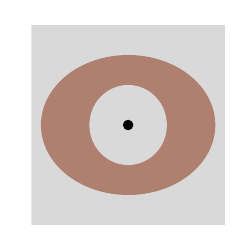}} & yes &    \ldots     &  Ulrich   & \ldots  & yes & variable & 5 & 10,000 \\
\texttt{s-ubsmi} & \parbox[c]{1.3cm}{\includegraphics[width=1.3cm]{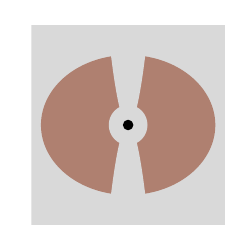}} & yes &    \ldots     &  Ulrich   & yes & yes & $\rsub$  & 7 & 10,000 \\
\texttt{s-ubhmi} & \parbox[c]{1.3cm}{\includegraphics[width=1.3cm]{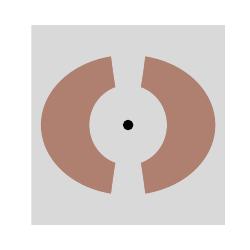}} & yes &    \ldots     &  Ulrich   & yes & yes & variable & 8 & 10,000 \\
\texttt{spu-smi} & \parbox[c]{1.3cm}{\includegraphics[width=1.3cm]{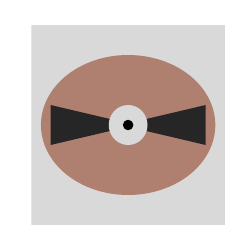}} & yes &  passive  &  Ulrich   & \ldots  & yes & $\rsub$  & 8 & 10,000 \\
\texttt{spu-hmi} & \parbox[c]{1.3cm}{\includegraphics[width=1.3cm]{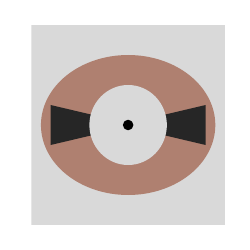}} & yes &  passive  &  Ulrich   & \ldots  & yes & variable & 9 & 10,000 \\
\texttt{spubsmi} & \parbox[c]{1.3cm}{\includegraphics[width=1.3cm]{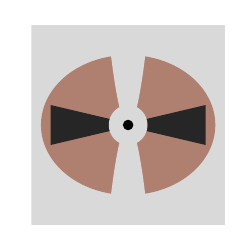}} & yes &  passive  &  Ulrich   & yes & yes & $\rsub$  & 11 & 40,000 \\
\texttt{spubhmi} & \parbox[c]{1.3cm}{\includegraphics[width=1.3cm]{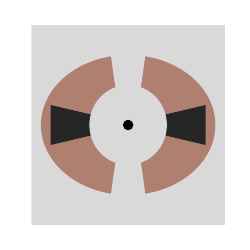}} & yes &  passive  &  Ulrich   & yes & yes & variable & 12 & 80,000 \\

\hline
\end{tabular}
\end{table*}

The other type of envelope included in the models is (as in R06) the
rotationally flattened envelope structure defined by \citet{ulrich:76:377}.
This envelope structure has a radial power-law dependence of around $3/2$
outside the centrifugal radius $\rc$, and $1/2$ inside. The density structure
has a singularity in the mid-plane at the radius $\rc$, which is due (in the
analytical model) to the pile-up of material due to conservation of angular
momentum. The density structure is given by

\begin{equation}
\rho(r,\theta) = \rho_0^{\rm env}\left(\frac{r}{\rc}\right)^{-3/2}\left(1 + \frac{\mu}{\mu_0}\right)^{-1/2}\left(\frac{\mu}{\mu_0} + \frac{2\mu_0^2\rc}{r}\right)^{-1}
,\end{equation}
where $\rho_0^{\rm env}$ is the scaling of the density structure which can be related (in the analytical model) to the infall rate $\mdote$:
\begin{equation}
\rho_0^{\rm env} = \frac{\mdote}{4\pi\left(G\mstar\rc^3\right)^{1/2}}
.\end{equation}
The $\mu_0$ value is given by the equation for the streamline:
\begin{equation}
\mu_0^3 + \mu_0\left(\frac{r}{\rc} - 1\right) - \mu\left(\frac{r}{\rc}\right) = 0
.\end{equation}
The envelope dust density is defined by specifying $\rho_0^{\rm env}$ rather than
$\mdote$ since the former had a direct impact on the SED and is not dependent
on the analytical model of infall (beyond the shape of the density structure
itself). The R06 models used this envelope by specified the envelope infall
rate $\mdote$, but this sometimes led to confusion since this infall is only used to set the density and does
not result in an increase in accretion luminosity. Users of the new models will still be able to
compute their own $\mdote$ values for the models, but by doing so manually, the
assumptions will be clearer.

For both envelope components, the envelope is truncated at the inner radius $\rmine$, and extends all the way to the edge of the grid, which is discussed in Section \ref{sec:methods}. The free parameters are the envelope density scaling $\rho_0^{\rm env}$, as well as the radial exponent $\gamma$ in the case of the power-law envelope, and the centrifugal radius $\rc$ in the case of the Ulrich envelope.

\subsubsection{Bipolar cavities}

Observations of embedded YSOs almost always show signs of bipolar outflows. The primary effect of these outflows on the circumstellar envelopes is to carve out
bipolar cavities by entraining the envelope material into the outflow. This
leads to a region of lower density where radiation can escape more easily, and
can sometimes be observed in scattered light (since the shorter wavelength
radiation finds a much easier escape route via the cavities).

Bipolar cavities can only be present in a model if an envelope
component is also present. The power-law cavity is defined as a region with
boundary given by
\begin{equation}
z(R) = z_0\left(\frac{R}{R_0}\right)^c
.\end{equation}
For $|z| > z(R)$, the dust density is set to $\rho_0^{\rm cav}$ or the envelope density, whichever is lowest. The value of $R_0$
and $z_0$ are set by defining the half-opening angle of the cavity $\theta_0$ at a spherical polar radius
of $r_0=10000$\,au:
\begin{align}
z_0 & = r_0\,\cos{\theta_0} ,\\
r_0 & = r_0\,\sin{\theta_0.}
\end{align}
The cavities are not empty, but instead contain dust with a constant density $\rho_0^{\rm cav}$, except in regions where the envelope density would be lower. The free parameters are the power-law exponent of the cavity opening $c$, the opening angle $\theta_0$, and the density inside the cavity $\rho_0^{\rm cav}$.

\subsubsection{Ambient medium}

\label{sec:ambient}

The last model component is the ambient interstellar medium.
Forming stars, especially embedded YSOs, are not completely isolated
objects, and the effect of the interstellar medium can sometimes be important.
As shown in \citet{whitney:13:30} and \citet{koepferl:14}, the interstellar
medium can be heated by the central stars and provide additional thermal
emission at the longest wavelengths.

For the models without an envelope, both models with and without an ambient
medium are included, while for models with envelopes the ambient medium is included in all cases.
The reason for always including it for envelopes is to avoid including a sharp cutoff in the envelope density,
since the analytical expressions for the envelope densities never go to zero. By adding
an ambient medium, the radiative transfer can be carried out to a radius
where both the density and the temperature reach a constant value.

The ambient medium, when included, is not simply a density that is added, but
instead it is a lower limit to the density (so that densities above it remain
unchanged). For the models presented here, the dust density was set to
$\rho_{\rm amb}=10^{-23}$\,g/cm$^3$, and the lower limit of the temperature in the
whole model was set to $T_{\rm amb}=10$\,K. As discussed in Section
\ref{sec:apertures}, we need to ensure that we correctly subtract the
background emission from this ambient medium once we extract the SEDs.

\subsection{Parameter space sampling}

\label{sec:sampling}

As mentioned in Section~\ref{sec:overall}, Table~\ref{table:parameters} lists the ranges of parameters sampled for each parameter. Table~\ref{table:grids} lists all the model sets computed for this initial model release. The model set name is indicated, along with the components that the model set includes. The model set name is composed of several characters that indicate which component is present. The characters, in order, are \texttt{s} (star), \texttt{p} (passive disk), \texttt{p} (power-law envelope) or \texttt{u} (Ulrich envelope), \texttt{b} (bipolar cavities), \texttt{h} (inner hole), \texttt{m} (ambient medium), and \texttt{i} (interstellar dust). If a component is absent, a `\texttt{-}' is given instead. Since these model names are not immediately intuitive, the second column shows an icon representing the basic layout of the model. In the figures shown in later sections, these icons are then used in figure panels to specify which model set is being shown.

All models include a star with the radius $\rstar$ and temperature $\tstar$ sampled from the ranges given in Table~\ref{table:parameters}. For the passive disks, power-law and Ulrich envelopes, and bipolar cavity components, the ranges sampled from are also given in Table~\ref{table:parameters}. For models where the ambient medium is included, the values are fixed to those given in Section~\ref{sec:ambient}.

For all models that contain at least a disk, envelope, or ambient medium, the inner radius is set to the same value for all components, and is either set to $\rsub$ (the dust sublimation radius) or is sampled from the range given in Table~\ref{table:parameters}.

The number of models in each set should ideally be related to the number of variable parameters in that set, but the simplest set has two parameters, while the most complex has 12, so the number of models is set using
\begin{equation}
n_{\rm models} = 10,000 \times
2^{{{\rm max}[n_{\rm var} - 9,0]}}
.\end{equation}

This resulted in 10,000 models for models with nine or fewer variable parameters. The reason for enforcing the lower limit on the number of models is to ensure that there were enough representative models for each set.

\subsection{Dust properties}

\label{sec:dust}

For this initial collection of models sets, the dust properties are taken to be the same for all models, specifically the dust properties from \citet{Draine:03:241, Draine:03:1017} using the \citet{Weingartner:01:296} Milky Way grain size distribution A for $R_{\rm V}$=5.5 and $C/H=30$\,ppm renormalized to $C/H=42.6$\,ppm. The full Mie scattering properties of this dust model were computed using the \texttt{bhmie} routine from \citep{Bohren:83}, modified by B. Draine\footnote{\url{http://www.astro.princeton.edu/~draine/scattering.html}}, and wrapped in a package that makes it easy to compute the properties for various size distributions\footnote{\url{https://github.com/hyperion-rt/bhmie}}.

As described in Section~\ref{sec:overall}, the aim of the set of models presented here is not to provide the most physically realistic models, but rather to provide a consistent set of models spanning a large region of parameter space, which can then be used to explore the behavior of various physical components in the models. While it might be desirable for example to vary the dust properties in disks to account for grain growth toward the mid-plane, the models provided here are a first step to ensure that in the first place a model with uniform dust properties cannot fit the observations. In many instances, one may be tempted to make models more complex rather than explore more of the existing parameter space.

In future, the set of models could of course be expanded in future to include models with different or even variable dust properties, and include for example the effects of polycyclic aromatic hydrocarbons (PAHs), but this initial set deliberately includes much simpler dust properties.

\section{Methods}

\label{sec:methods}

\subsection{Radiative transfer code}

The code used to compute the models presented here is \textsc{Hyperion}, an open-source 3-d Monte-Carlo dust continuum radiative transfer code, described in detail in \citet{robitaille:11:A79}. The models presented here were computed with version \texttt{v0.9.5} of the code.

The radiative transfer was computed on a spherical polar grid. For models with spherical symmetry, a 1-d grid with 400 radial cells was used, while for more complex axisymmetric models, a 2-d grid with 400 radial cells and 300 polar cells was used. The radial cells were distributed so as to ensure that the width of a cell at the inner edge of the disk was small enough to resolve the temperature gradient well. This was done by picking values equally spaced in log space between $\delta R$ and $\rmaxd - \rmind$, where $\delta R$ is the width at the inner edge of the disk corresponding to $\tau=0.1$ at the peak wavelength of the stellar spectrum, then adding these values to $\rmind$ (this is the default mode in \textsc{Hyperion} when computing YSO models). The outer radius of the grid was set to $\sqrt{2}$ times the radius of the largest aperture in which the SEDs were measured. The aperture sizes as well as the origin of the $\sqrt{2}$ factor are discussed in Section \ref{sec:apertures}.

Rather than using dust temperatures, the code carries out all calculations in terms of the absorbed energy per unit mass of dust (the absorbed specific energy). In the first few iterations, the code computes the absorbed specific energy in each cell using the algorithm outlined by \citet{lucy:99:282}. Once the code detects that the calculation has converged, it then proceeds to computing the SEDs. The convergence detection is described in detail in \citet{robitaille:11:A79} -- for the models presented here, the specific energy is considered to be converged once the 99th percentile value of the energy difference is less than a factor of two, and the change in this value changes less than 10\% from one iteration to the next.

The modified random walk (MRW) algorithm \citep{min:09:155, robitaille:10:A70} was used in order to speed up the computation of the Monte-Carlo propagation in optically thick regions. The raytracing algorithm described in \citet{robitaille:11:A79} was also used in order to obtain excellent S/N at long wavelengths, to avoid issues such as those in the R06 models (c.f. \S\ref{sec:lowsn}).

\subsection{Spectral energy distributions}

\label{sec:seds}

\subsubsection{Overview}

The SEDs were computed for 250 wavelength bins logarithmically spaced between 0.01\microns and 5\,mm. The SEDs also contain information about the origin of the emission, for example whether it comes directly from the star, whether it was scattered, or whether it was emitted by dust. Each SED was computed for all four Stokes parameters I, Q, U, and V, which means that in addition to the SED of the total flux, polarization spectra are also available. The linear polarization is given by
\begin{equation}
p_{\rm lin} = \sqrt{\frac{Q^2 + U^2}{I^2}}
,\end{equation}
and the circular polarization by
\begin{equation}
p_{\rm circ} = \left|\frac{V}{I}\right|
.\end{equation}

\subsubsection{Viewing angles}

The SEDs were computed for nine viewing angles for each model. In contrast to the R06 models, where the SEDs were produced using escaping photons binned into viewing angle ranges, and thus effectively producing SEDs averaged over a range of viewing angles, the models presented here were computed using peeling-off and raytracing \citep[c.f.][]{robitaille:11:A79}, which allows one to compute SEDs at exact viewing angles. However, selecting the same viewing angles for all models is not ideal, since it does not adequately sample the full range of viewing angles and favors specific ones. For example, if SEDs are computed from $0^\circ$ to $90^\circ$ in steps of $10^\circ$, then model SEDs for disks will vary dramatically between $80^\circ$ and $90^\circ$ (edge-on), and there will be no SEDs for viewing angles in between these values, so a source seen at a viewing angle of $85^\circ$ will not have any adequate models. Therefore, each model was instead computed for 9 viewing angles randomly sampled between $0^\circ$ to $90^\circ$. However, since when looking at a model it is useful to see how the SED changes with viewing angle, and we do not want models to randomly have all angles close to pole-on, I used stratified sampling, which means randomly sampling one angle between $0^\circ$ and $10^\circ$, one between $10^\circ$ and $20^\circ$, and so on, which still produces a distribution that is random for our purposes.

\subsubsection{Apertures and extent along the line of sight}

\label{sec:apertures}

The SEDs were computed in 20 circular apertures, with radii logarithmically spaced between the dust sublimation radius and an outer radius that depended on the type of model. For models with no ambient medium but with a disk, the largest aperture was simply set to the radius of the disk. For the star-only models with no ambient medium, all apertures were set to include all the flux from the star.

For models with an ambient medium, the situation required a more careful treatment of the outer aperture as well as the emission region in three dimensions. Let us first consider the example of a constant density optically thin ambient medium with no central source: in this case, if the ambient density is set up in a spherical polar grid, then an observer would not observe a constant surface brightness, but rather would see a circularly symmetric region of emission peaking at the center and falling off toward the edges (essentially the projection of an optically thin sphere onto the sky). However, this is not desirable, since we need a model with a constant density and temperature to produce a constant surface brightness on the sky.

\begin{figure}[t]
\begin{center}
\includegraphics[width=9cm]{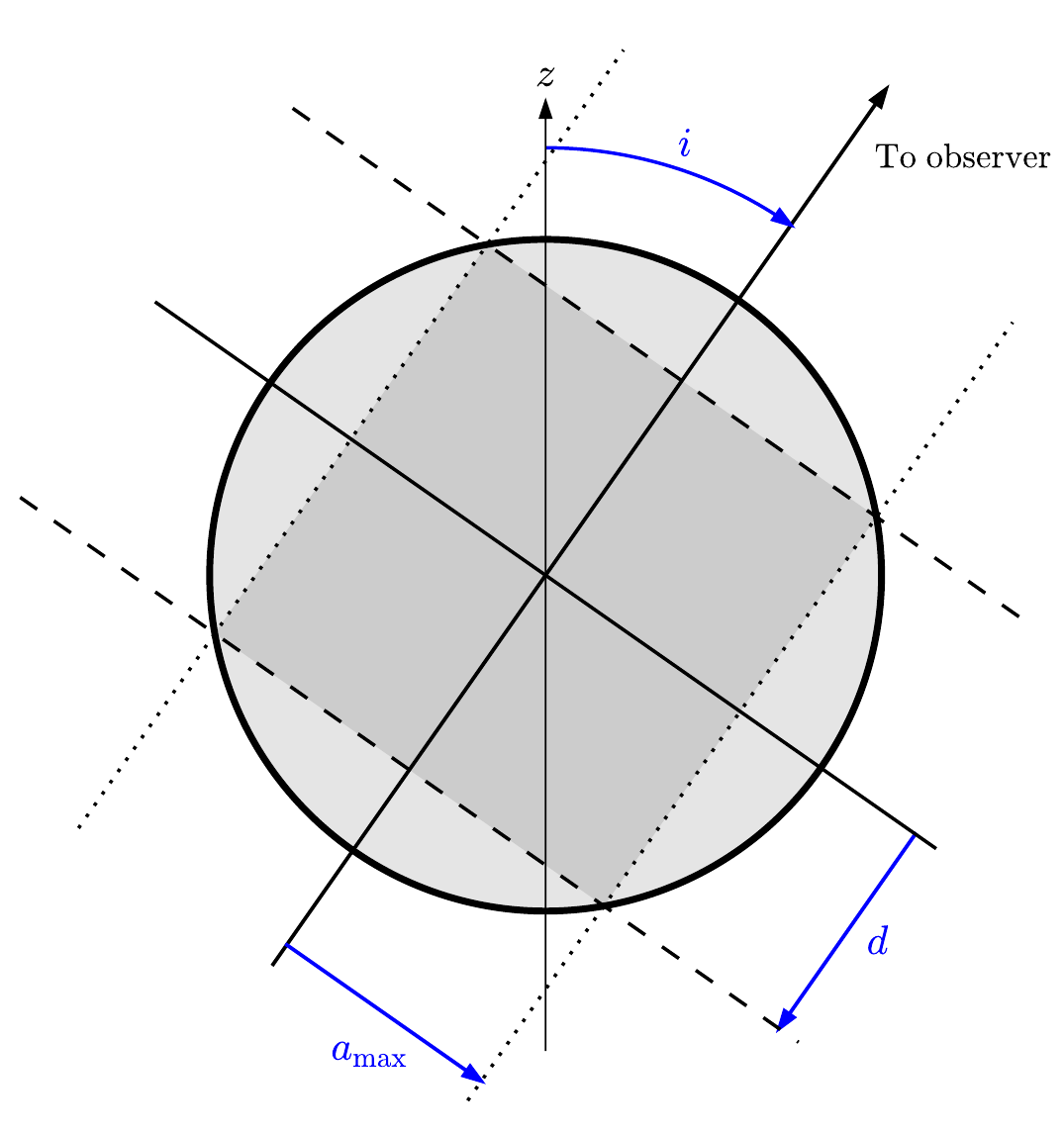}
\caption{Schematic representation of a typical model. The $z$ axis of the model is shown as the black vertical arrow. In this example, the line of sight is rotated by an inclination of $i=35^\circ$ relative to the $z$ axis. The dotted lines show the extent of the largest aperture with radius $a_{\rm max}$, while the dashed lines show the extent of the slab with half-width $d$ inside which photons are taken into account for the SED. To ensure that the models account for all the emission in the intersection of the largest aperture and the slab (the dark shaded area), the edge of the grid (the thick solid circle) needs to have a radius of $\sqrt{2}d$. \label{fig:diagram}}
\label{default}
\end{center}
\end{figure}

Hyperion allows images and SEDs to be computed by including only photons emitted in a fixed range of distances along the line of sight - thus, one can in effect set up a slab outside which photons are ignored. This slab is set up to have a half-width $d$ that is the same as the radius of the largest aperture $a_{\rm max}$. Thus, the region of emission included in the SEDs is therefore a cylinder rather than a sphere, which avoids the issues mentioned above. In order to make sure that the cylinder is completely filled, the spherical grid itself thus needs to extend to $\sqrt{2}$ times the radius of the largest aperture (also the half-width of the slab). A diagram of this set-up is shown in Figure \ref{fig:diagram}.

With this set-up, as the apertures become larger, the surface brightness should asymptotically tend to the surface brightness of the ambient medium. The transition point at which the surface brightness reaches this level is $r(\rho=\rho_{\rm amb})$ or $r(T=T_{\rm amb})$, whichever is largest. While $r(\rho=\rho_{\rm amb})$ is known from the start, it is impossible to know $r(T=T_{\rm amb})$ in advance of running a model, since the temperature structure is required. One could therefore choose the radius at which the temperature would reach the ambient level in the optically thin limit. However, for optically thick models, since the temperature could in reality approach the ambient level at a much smaller radius, we would be wasting many photon packets to the constant density ambient region. Therefore, the radiative transfer was first run to determine the temperature structure, and only then were the final apertures to use for the SED decided.

\subsubsection{Postprocessing}

Once the SEDs were computed, a background surface brightness was subtracted from each aperture. This background surface brightness was determined by computing the modified blackbody emission that would arise from a medium with density $\rho_{\rm amb}$, temperature $T_{\rm amb}$, thickness $2d$, and assuming the dust properties described in Section \ref{sec:dust}. To avoid S/N issues, any flux densities for smaller apertures that are below 0.1\% of the SED in the largest aperture are masked (ignored), for wavelengths larger than 10\microns \,-- this was done because as the apertures get smaller, the long-wavelength emission becomes dominated by a smaller number of photons, and under this threshold the flux densities start to become unreliable. Finally, the model flux densities were all interpolated to a common set of 20 apertures logarithmically spaced between $10^2$au and $10^6$au.

\subsection{SED fitting code}

\label{sec:fitter}

In R07, we presented a fitting algorithm for the R06 models, which was implemented as a Fortran fitting code that could be called from a public web interface. The underlying Fortran code was, however, not publicly available. While the web interface allowed for easy modeling of low numbers of sources, it was not suitable for modeling many forming stars across entire regions. I have now written a new Python-based fitting code that allows users to carry out modeling of many sources using the new models (as well as the R06 models). This will also allow users to customize the fitting workflow, and for example to eliminate unphysical models, or fit models from the different sets to the same data. The new code is available\footnote{\url{http://doi.org/10.5281/zenodo.235786}} under an open-source BSD license, and was used for the results shown in Section~\ref{sec:fitting_examples}.

\section{Results}

\label{sec:results}

\subsection{Computational details}

\label{sec:computational}

The models were computed on a cluster with 120 cores over a period of several months.  The total CPU time to compute the models was approximately 640,000 hours. A small fraction (just below 2\%) of models were terminated during the temperature calculations since they did not complete within the allocated time due to extremely high optical depths, and they are therefore missing from the initial \texttt{v1.1} release of the models. However, they will be added to future releases.

In addition, for the two sets of models with spherically symmetric power-law envelopes, the temperature calculations for some of the models did complete, but the calculation of the SEDs did not. For the \texttt{v1.1} release of the models, the SEDs for these models were therefore recomputed without including the contribution from scattered stellar and scattered dust emission. Since this concerns the models with the highest optical depth, not including the scattered stellar emission is a valid approximation. On the other hand, for a subset of the models, not including the scattered dust emission may have an effect. By extrapolating the relationship between the contribution of the scattered dust emission to the total SED as a function of wavelength, it is possible to infer that between 2 and 14\microns, for 90\% of the models, the effect of not including the scattered dust emission will result in a less than 10\% effect. Outside this wavelength range, the effect will not be noticeable for any of the models. In future, Hyperion will include a much more efficient method for computing the scattered light.

\subsection{Models}

\begin{figure*}[htbp]
\begin{center}
\includegraphics[width=18cm]{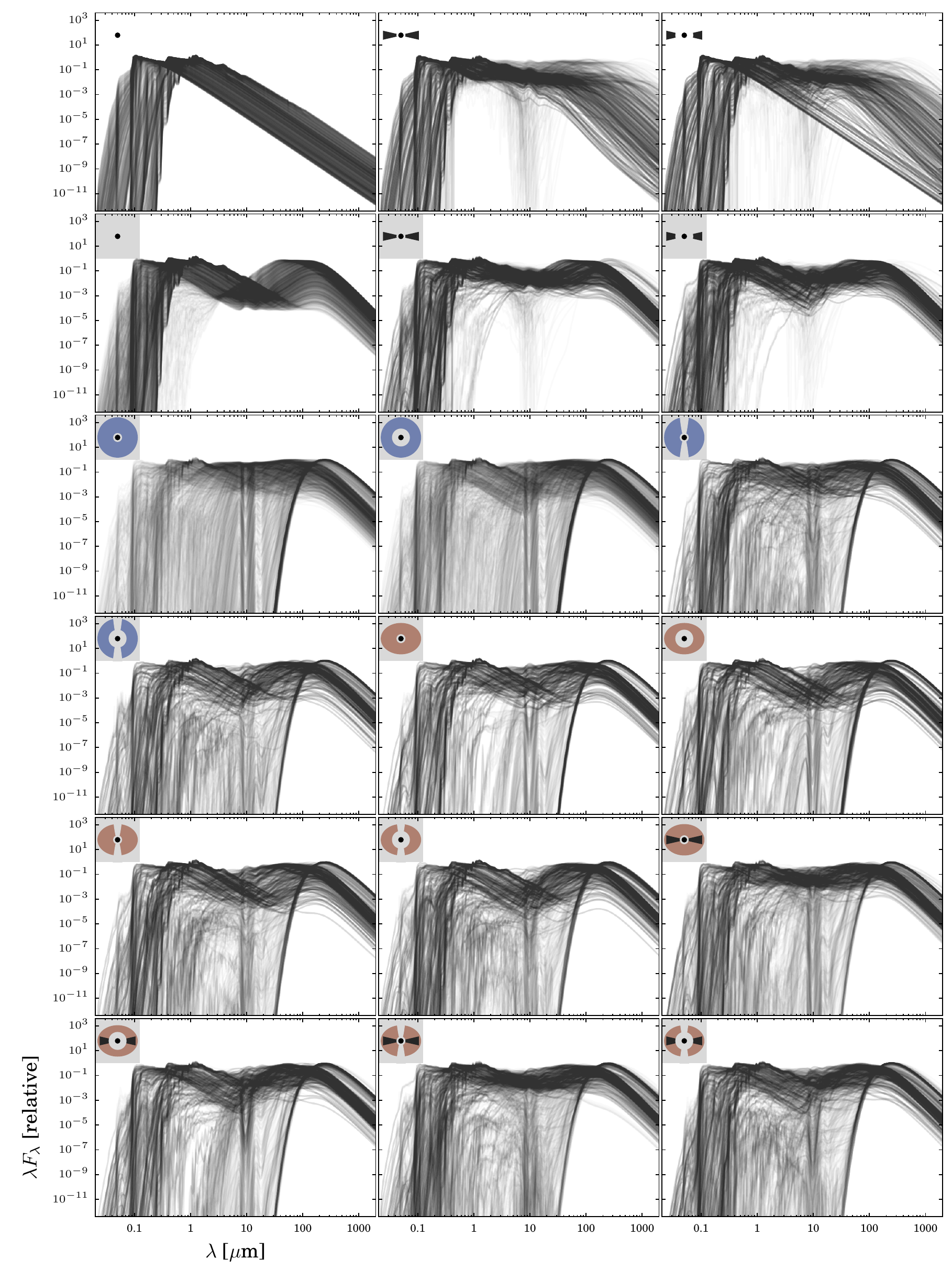}
\caption{A subset of 2000 SEDs for each model set, normalized to the total luminosity of each SED. \label{fig:all_seds}}
\label{default}
\end{center}
\end{figure*}

Figure~\ref{fig:all_seds} shows an overview of the SEDs in the model sets by selecting 2000 SEDs for each model set. These are normalized to the total luminosity of the SED to make it easier to compare them to one another. The aim of this figure is not to show each individual SED but to show trends over the different model sets. For example, as expected, models with envelopes have a much wider range of near- and mid-infrared fluxes compared to models with a disk alone.

\subsection{Model availability}

\label{sec:avaiability}

The models are publicly available\footnote{\url{http://doi.org/10.5281/zenodo.166732}}. Given the number of models, these are primarily delivered using a few large files, rather than millions of small files. Each model set consists of the following main files:

\begin{itemize}
\item a FITS file containing the fluxes and associated uncertainties for all models, apertures, and wavelengths
\item a FITS file containing the parameters for all the models
\end{itemize}

The motivation for providing the models as a few large files is that it allows the data to be more easily sliced, for example to return the flux at a specific wavelength for all models, for the largest aperture, or the fluxes for all apertures and wavelengths for a given model, and so on. Several FITS file readers such as that included in Astropy \citep{astropy:13} are able to read subsets of data from FITS files as needed, meaning that the data can still be accessed if the FITS files are larger than the available memory.

In addition to these main files, I provide the model fluxes obtained by convolving the SEDs with a number of common transmission curves, for example for instruments on \textit{Spitzer} or \textit{Herschel}, and I also provide a utility in the SED fitting tool (c.f. \S\ref{sec:fitter}) that allows users to convolve the models with any other transmission curve. Finally, I am also making available the individual model files, which can be used to extract information relating to the density and temperature grids, as well as the polarization spectra.

Rather than consider the models a static and final version, the models are versioned and may be updated in future if bugs or other issues are found at any stage of the pipeline to generate the models. The version used in this paper is \texttt{v1.1}, and this will be updated if any issues are fixed.

\subsection{Fitting strategy}

\label{sec:fitting_examples}

\subsubsection{General approach}

In R07 and subsequent studies, the approach taken to modeling observations with the R06 models was to simply fit all the models to the observations and directly look at the parameter values in order to place constraints on the physical properties.
The model sets presented here allow us to better separate the problem into determining constraints related to the presence or absence of components (e.g., `Is there any evidence for a disk in this source?'), and determining constraints related to the physical parameter values (e.g., `What is the disk mass?').

As mentioned in Section~\ref{sec:overall}, one of the consequences of the uniform sampling of parameter space is that not all models are necessarily physically realistic, so when fitting real objects, the user of the models is responsible for deciding which models are realistic. However, for the example presented in this section, all models are assumed to be equally realistic.

\subsubsection{Data and fitting parameters}

In this section, I show an example of modeling observational data using the new models. The data used is that for a source in the NGC2264 star formation region \citep{forbrich:10:1453}. The source, referred to as Source 20 in \citeauthor{forbrich:10:1453}, is the 2MASS source J06404862+0935578 at a position of $\alpha=06^h40^m48.62^s\,\,\delta=09^\circ35^\prime57.8^{\prime\prime}$ (ICRS). The fluxes extend from optical to sub-millimeter wavelengths, and were carefully checked as part of that study. In cases where the source was undetected, upper limits are provided.

As in \citet{forbrich:10:1453}, the distance range was assumed to be 869 to 961\,pc (within $\sim5$\% of $d=913$\,pc), and A$_{\rm V}$ range to be from 0 to 40\,mag. The extinction law used was the same as that used in \citeauthor{forbrich:10:1453}

\subsubsection{Results}

In Figure \ref{fig:seds_20_rel} I show for each model set presented here all the models that satisfy $\chi^2 - \chi^2_{\rm best} < 3\,n_{\rm data}$, where $\chi^2_{\rm best}$ is the $\chi^2$ of the best model for each model set (see R07 for details on the $\chi^2$ calculation). The panels in the figure are ordered roughly in order of increasing complexity from left to right, then from top to bottom, so that the top left panel shows models that include only a star, and the bottom right panel shows models that include a star, disk, infalling envelope, bipolar cavity, and an ambient medium. At a first glance, a substantial fraction of the model sets provide a good fit to the data, with the exception of the models with neither disk or envelope, and the models with only an Ulrich envelope. All model sets with only a disk or only a power-law envelope, or with an envelope and a disk provide a good fit.

It thus initially appears that one cannot place strong constraints on the nature of the object. However, I now show that one can use information about not only the quality of the fit, but also the fraction of models that provide a good fit in order to try and determine which model is most likely (even though many of the models remain possible).

\subsubsection{Bayesian model comparison}

\label{sec:bayesian}

Using a Bayesian approach, we can compare how well two models $M_1$ and $M_2$ explain a set of data $D$ using the Bayes factor

\begin{equation}
K \equiv \frac{P(D|M_1)}{P(D|M_2)}
.\end{equation}

In this equation, the $P(D|M_i)$ terms are the likelihoods, which represent the probability that the data $D$ are produced if the model $M_i$ is correct. Thus, the Bayes factor gives the ratio of the probability that one model produces the data compared to the other. If $K=1$, then the two models are equally likely, whereas if $K>1$ then model $M_1$ is more likely than $M_2$ (and vice-versa).

Assuming that a model $M$ can be parametrized by a parameter vector $\vec{\theta}$, we can expand the likelihood as an integral over parameter space:

\begin{equation}
P(D|M) = \int p(\vec{\theta}|M)\,p(D|\vec{\theta},M)\,d\vec{\theta}
.\end{equation}

In our case, a model $M$ corresponds to one of our model sets with a fixed number of parameters. For each model $M$ we  have a finite number of pre-existing samples of parameter space. We can therefore approximate $P(D|M)$ by a sum over the existing model samples $\vec{\theta_{j}}$:
\begin{equation}
P(D|M) = \sum_{j=1}^{N}P(\vec{\theta_{j}}|M)\,P(D|\vec{\theta_{j}},M)
,\end{equation}
where $N$ is the number of samples. Assuming that the models have been sampled such that all models samples are equally likely, the term $P(\vec{\theta_{j}}|M)$ is then equal to $1/N$, and we can re-write the likelihood as

\begin{equation}
\label{eq:final_likelihood}
P(D|M) = \frac{1}{N}\,\sum_{j=1}^{N}P(D|\vec{\theta_{j}},M)
.\end{equation}

In other words, the likelihood is the sum of the probabilities of that each model sample will produce the data, divided by the number of samples, or more simply, the likelihood is simply the probability of reproducing the data averaged over the model samples. Comparing which model is most likely is then simply a matter of comparing the values of $P(D|M)$ for different models $M$. We note that the assumption about all model samples being equally likely does mean that we are treating the initial sampling of the models (i.e., logarithmic or linear) as a prior. It is also important to note that uniform (linear or logarithmic) sampling does increase the risk of missing small areas of high probability, and the models presented here should be considered as a starting point for any modeling effort rather than a set that includes the optimal model for any specific dataset.

The implication of Eq. (\ref{eq:final_likelihood}) is that given two different model sets, even if both can fit the data equally well, the one where the highest fraction of models fit well should be preferred. If one considers a simple model set, for example a model set with only a star and a disk, and then compares this to a model set with an envelope added, the region of parameter space providing a good fit in the model with the envelope will represent a smaller fraction of the total parameter space. Therefore, if both model sets have the same total number of samples, the region of parameter space providing a good fit will be sampled by fewer models, and therefore the fraction of models providing a good fit will be lower. Thus, the Bayes factor is simply a formal mathematical way of thinking about Occam's razor, which states that all things considered, the simpler model should be favored because more complex models require more fine tuning to reproduce the data.

The challenge in our case is to estimate the probability $P(D|\vec{\theta_{j}},M)$. Formally, under the assumption of normal errors, we can write this as

\begin{equation}
P(D|\vec{\theta_{j}},M) \propto \exp{\left(-\frac{\chi^2}{2}\right)}
.\end{equation}

\begin{figure*}[htbp]
\begin{center}
\includegraphics[width=18cm]{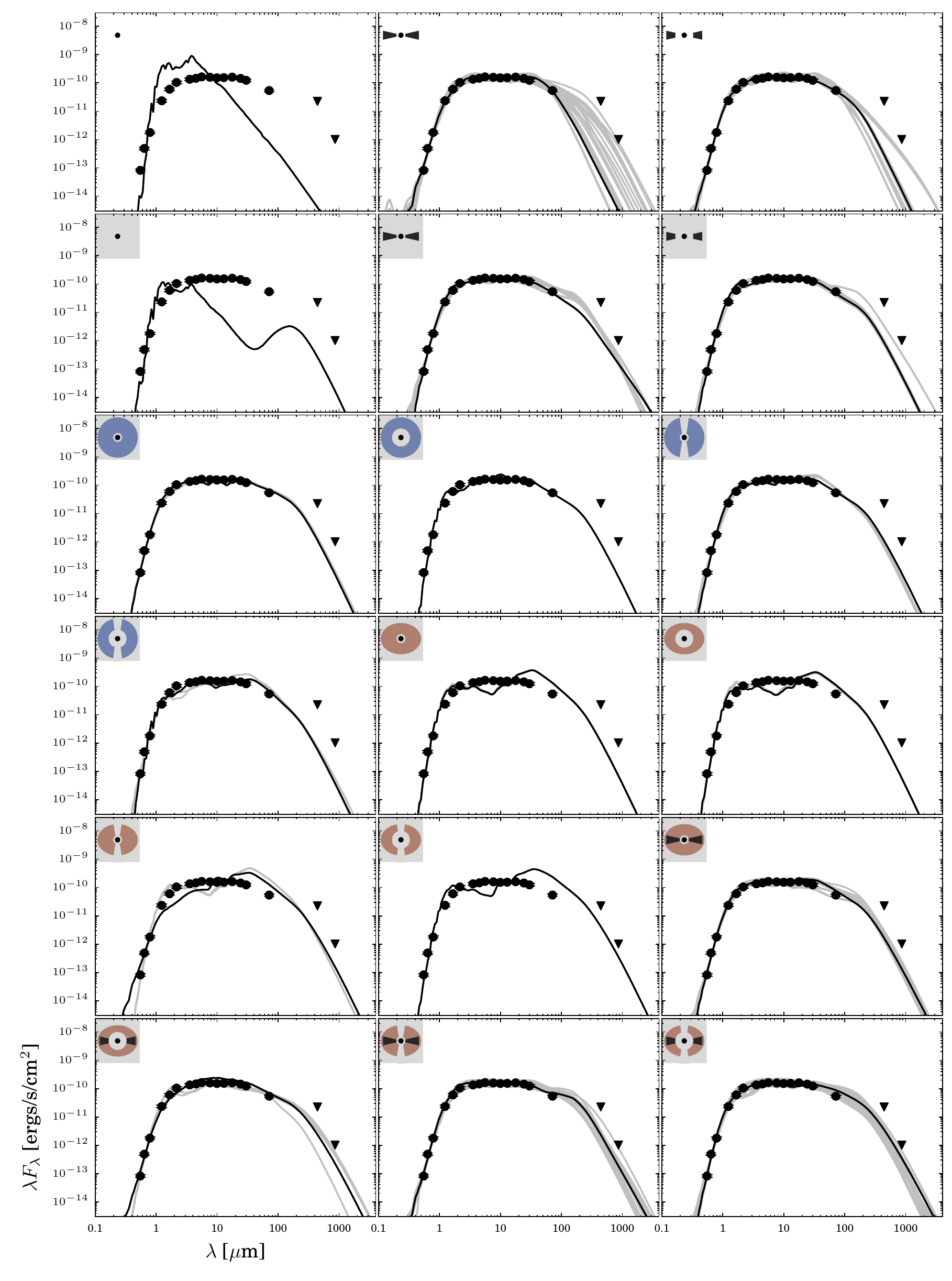}
\caption{Model fits to observations of source 20 in NGC2264 (using the source nomenclature from \citealt{forbrich:10:1453}), where each panel shows all the fits for $\chi^2 - \chi^2_{\rm best} < 3\,n_{\rm data}$ for a specific model set (and where $\chi^2_{\rm best}$ is determined for each model set individually).\label{fig:seds_20_rel}}
\end{center}
\end{figure*}

However, as I will discuss in Section \ref{sec:misspecification}, this is often too stringent a definition, because in the case of observed SEDs, there are  a number of systematic sources of errors to consider, such as variability or the fact that the models are only simplistic representations of a more complex 3-d reality. Therefore, I suggest here (as was done in R07) that we should adopt a less stringent definition of probability. For example, in R07, the approach was to treat all models with
\begin{equation}
\chi^2 - \chi^2_{\rm best} < 3\,n_{\rm data}
,\end{equation}
as being `good', and remaining models as being `bad'. Of course, this is a very binary definition that also does not reflect the continuum of good to bad fits, but it corresponds roughly to what one would consider reasonable vs. unreasonable models by eye. If we adopted this definition, we would be treating $P(D|\vec{\theta_{j}},M)$ as

\begin{equation}
\label{eq:simple_prob}
P(D|\vec{\theta_{j}},M) \propto \left\{\begin{array}{ccc}1 & \rm{for} & \chi^2 - \chi^2_{\rm best} \le 3\,n_{\rm data} \\0 & \rm{for} & \chi^2 - \chi^2_{\rm best} > 3\,n_{\rm data}\end{array}\right.
.\end{equation}

Yet another possibility would be to correctly model all the systematic effects in SED modeling, and determine a real $P(D|\vec{\theta_{j}},M)$, although in many cases this is not possible without significant effort. The aim here is not to debate which definition of $P(D|\vec{\theta_{j}},M)$ should be used, and this will depend on the problem being solved.

We can now consider the example from Figure~\ref{fig:seds_20_rel} and try to apply this formalism in order to determine which model set is the most likely. We adopt a definition of the probability from Eq.~(\ref{eq:simple_prob}), but in order for the comparison to be fair between different models, we need to use the same value of $\chi^2_{\rm best}$ for all models. Otherwise, a model set that has many poor models, where the best fit is poor, may rank the same as a model set that has many good models, where the best fit is good. We therefore define $\chi^2_{\rm best}$ as being the $\chi^2$ for the single best-fitting model over all model sets. For the purpose of this example, a criterion of
\begin{equation}
\label{eq:abs_criterion}
\chi^2 - \chi^2_{\rm best} \le 9\,n_{\rm data}
,\end{equation}
provided a sensible separation between good and bad fits, so models that satisfy this are assigned a probability of one, while the remaining ones have probability of zero. If we write the number of `good' models as $N^{\rm good}$ for the model set $M$, we can then re-write Eq.~(\ref{eq:final_likelihood}) as

\begin{equation}
\label{eq:final_likelihood2}
P(D|M) \propto \frac{N^{\rm good}}{N}
.\end{equation}

In other words, we simply look at which model set has the highest fraction of good models. In this way, even if a specific model set contains the best fit by $\chi^2$ value, if only one model provides a good fit in that set, it means that the parameters need to be fine-tuned to reproduce the data, while a model set where a larger fraction of models can reproduce the data is more likely because it requires less fine-tuning.

\begin{figure*}[htbp]
\begin{center}
\includegraphics[width=18cm]{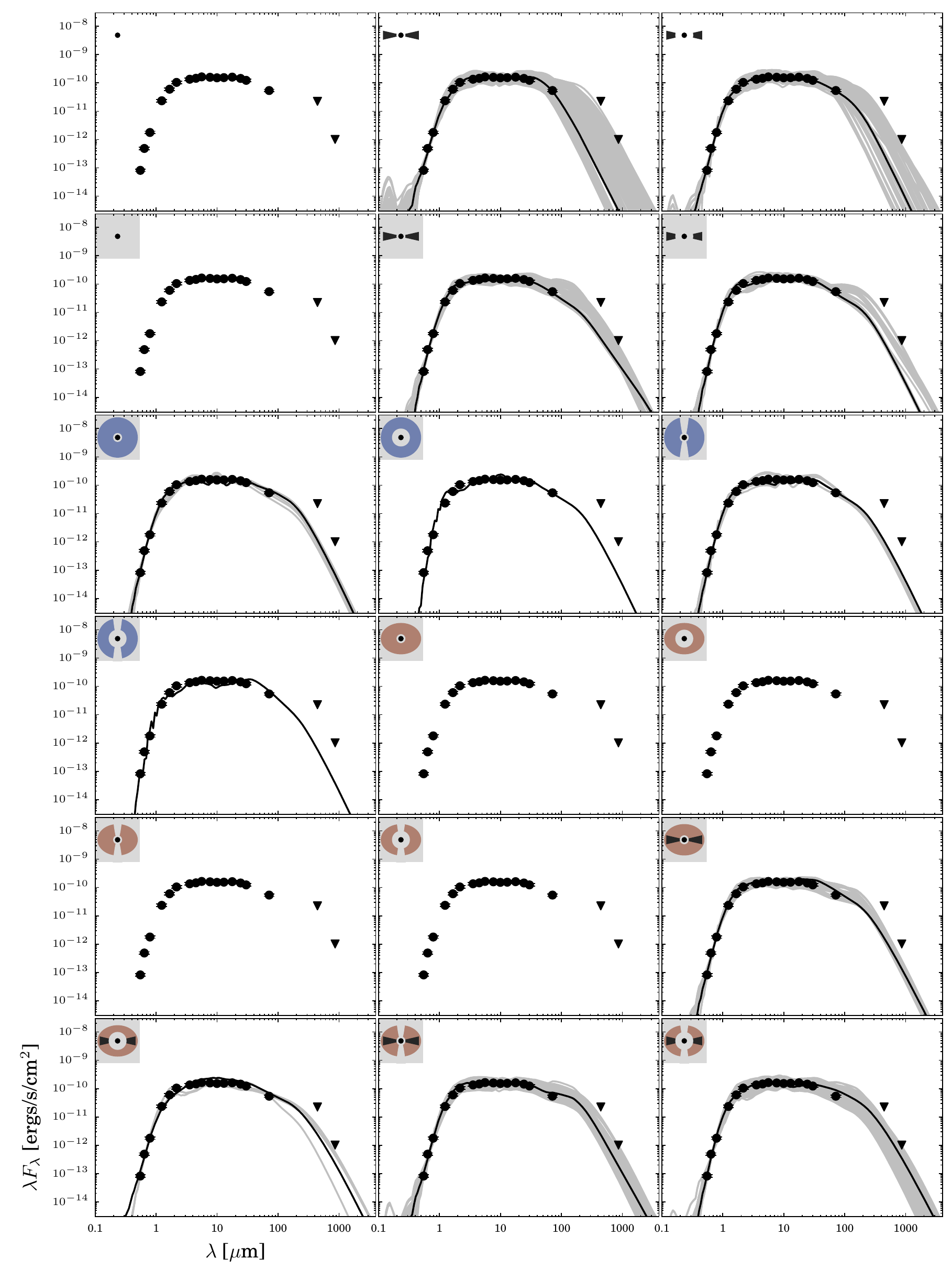}
\caption{As in Figure \ref{fig:seds_20_rel}, but showing all the fits for $\chi^2 - \chi^2_{0, \rm best} < 9\,n_{\rm data}$ where $\chi^2_{0, \rm best}$ is the same for all panels, and set to the best fit over all model sets.\label{fig:seds_20_abs}}
\end{center}
\end{figure*}

Figure~\ref{fig:seds_20_abs} shows all the models considered good when using the same $\chi^2_{\rm best}$ for all models. This shows that already some model sets cannot provide any good fits to the data according to this definition. We now look at the values of $P(D|M)$ for the remaining model sets.

Table~\ref{tab:likelihoods} lists for each model set the fraction of models providing a good fit, the best-fit $\chi^2$ value, and the relative score, which is given by the ratio of $P(D|M)$ to the mean of the $P(D|M)$ values for all model sets. A higher relative score means a more likely model. Using this, we can see that the most likely model is that of a simple disk with no additional inner hole, no envelope, and no ambient medium. If we look at the model set that also includes an ambient medium, the $\chi^2$ value for this model set is lower (2.28 compared to 2.71), but because a smaller fraction of models fit, the model is overall less likely (the score is $1.947$ compared to $7.600$ for the model with no ambient medium). Similarly, the model with a flattened envelope, bipolar cavities, and embedded in an ambient medium has a best-fit $\chi^2$ value only slightly larger ($3.00$ instead of $2.71$), but the score is much lower (0.867 vs 7.600). This is because the model with the envelope requires more fine-tuned conditions in order to reproduce the data. Of course, the model with the envelope may actually be correct, but the Bayes factor tells us that without any other information, it is easier for the simpler model to reproduce the data, and we should therefore consider it more likely.

\begin{table}[htp]
\caption{Relative likelihoods of the model sets for the example source modeled\label{tab:likelihoods}}
\begin{center}
\begin{tabular}{|l|c|ccc|}
\hline
 &  & $P(D|M_i)$ & & Score\\
Model set & Icon & (relative) & $\chi^2_{\rm best}$ & (relative)\\
\hline
\texttt{s---s-i} & \parbox[c]{1.3cm}{\includegraphics[width=1.3cm]{icon_s---s-i.pdf}} & 0.0000\% & 384.60 & 0.000 \\
\texttt{sp--s-i} & \parbox[c]{1.3cm}{\includegraphics[width=1.3cm]{icon_sp--s-i.pdf}} & 0.3167\% &   2.71 & \textbf{7.600} \\
\texttt{sp--h-i} & \parbox[c]{1.3cm}{\includegraphics[width=1.3cm]{icon_sp--h-i.pdf}} & 0.1156\% &   2.71 & 2.773 \\
\texttt{s---smi} & \parbox[c]{1.3cm}{\includegraphics[width=1.3cm]{icon_s---smi.pdf}} & 0.0000\% & 713.92 & 0.000 \\
\texttt{sp--smi} & \parbox[c]{1.3cm}{\includegraphics[width=1.3cm]{icon_sp--smi.pdf}} & 0.0811\% &   2.28 & 1.947 \\
\texttt{sp--hmi} & \parbox[c]{1.3cm}{\includegraphics[width=1.3cm]{icon_sp--hmi.pdf}} & 0.0267\% &   3.38 & 0.640 \\
\texttt{s-p-smi} & \parbox[c]{1.3cm}{\includegraphics[width=1.3cm]{icon_s-p-smi.pdf}} & 0.0700\% &   3.99 & 1.680 \\
\texttt{s-p-hmi} & \parbox[c]{1.3cm}{\includegraphics[width=1.3cm]{icon_s-p-hmi.pdf}} & 0.0100\% &  11.12 & 0.240 \\
\texttt{s-pbsmi} & \parbox[c]{1.3cm}{\includegraphics[width=1.3cm]{icon_s-pbsmi.pdf}} & 0.0178\% &   3.50 & 0.427 \\
\texttt{s-pbhmi} & \parbox[c]{1.3cm}{\includegraphics[width=1.3cm]{icon_s-pbhmi.pdf}} & 0.0011\% &  10.42 & 0.027 \\
\texttt{s-u-smi} & \parbox[c]{1.3cm}{\includegraphics[width=1.3cm]{icon_s-u-smi.pdf}} & 0.0000\% &  40.80 & 0.000 \\
\texttt{s-u-hmi} & \parbox[c]{1.3cm}{\includegraphics[width=1.3cm]{icon_s-u-hmi.pdf}} & 0.0000\% &  39.24 & 0.000 \\
\texttt{s-ubsmi} & \parbox[c]{1.3cm}{\includegraphics[width=1.3cm]{icon_s-ubsmi.pdf}} & 0.0000\% &  46.59 & 0.000 \\
\texttt{s-ubhmi} & \parbox[c]{1.3cm}{\includegraphics[width=1.3cm]{icon_s-ubhmi.pdf}} & 0.0000\% &  56.08 & 0.000 \\
\texttt{spu-smi} & \parbox[c]{1.3cm}{\includegraphics[width=1.3cm]{icon_spu-smi.pdf}} & 0.0489\% &   4.09 & 1.173 \\
\texttt{spu-hmi} & \parbox[c]{1.3cm}{\includegraphics[width=1.3cm]{icon_spu-hmi.pdf}} & 0.0056\% &   8.51 & 0.133 \\
\texttt{spubsmi} & \parbox[c]{1.3cm}{\includegraphics[width=1.3cm]{icon_spubsmi.pdf}} & 0.0361\% &   3.00 & 0.867 \\
\texttt{spubhmi} & \parbox[c]{1.3cm}{\includegraphics[width=1.3cm]{icon_spubhmi.pdf}} & 0.0206\% &   3.64 & 0.493 \\

\hline
\end{tabular}
\end{center}
\end{table}

\subsubsection{Parameter space}

\begin{figure*}[htbp]
\begin{center}
\includegraphics[width=6cm]{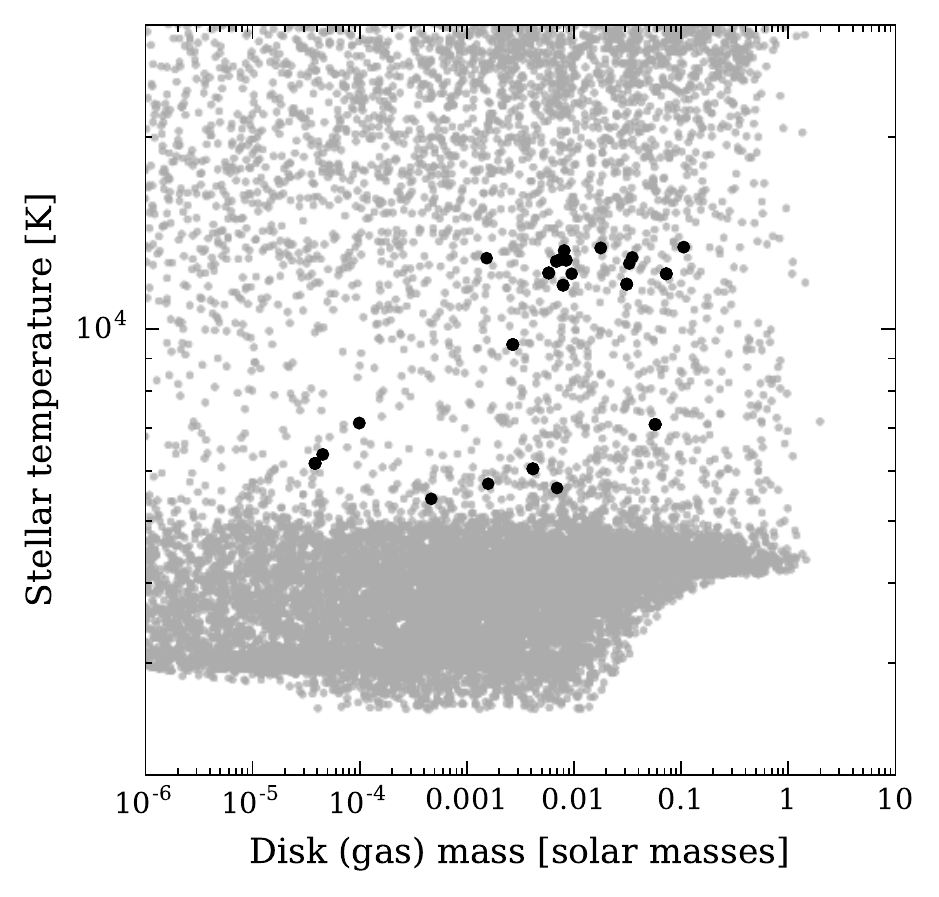}
\includegraphics[width=6cm]{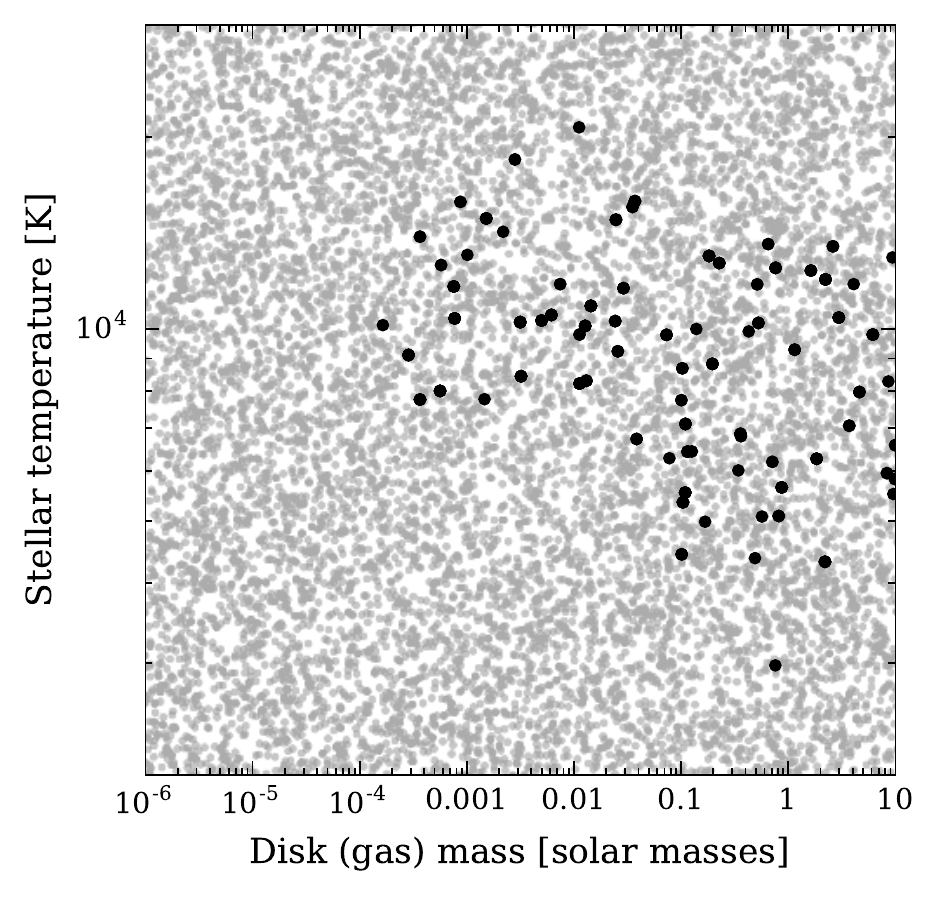}
\includegraphics[width=6cm]{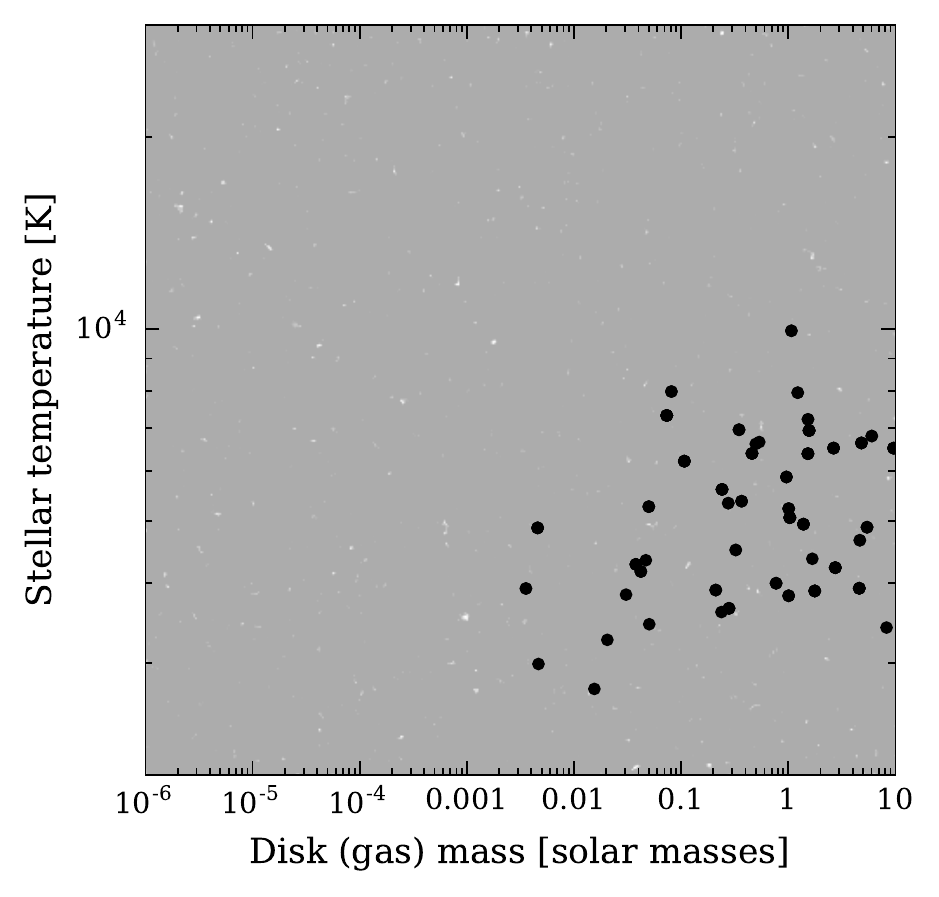}
\caption{Comparison of parameter space coverage of all the models (gray points) and best fits to the source shown in Figure~\ref{fig:seds_20_abs} (black) for the R06 model grid (left), the subset of models presented in this paper that are in the \texttt{sp--s-i} set (center), and the \texttt{spubhmi} set (right). The good fits are defined using Eq. (\ref{eq:abs_criterion}), and we apply the same criterion to the R06 models. The limits are adjusted to those covered by the new model sets, and the left figure would include more points on the left and higher temperatures if the plot limits were expanded. While the disk masses sampled for the new models were dust masses, these have been converted to gas masses for the above figures assuming a gas-to-dust ratio of 100.\label{fig:results_par}}
\end{center}
\end{figure*}

Having identified a most likely model set, we can take a look at some of the parameters from the models that fit well. However, since other model sets provided reasonable fits, we examine the parameters only under the assumption that this most likely model set is correct. Figure~\ref{fig:results_par} shows a 2-d projection of parameter space, showing the stellar temperature against the total disk mass. The figure compares the results obtained with the R06 grid versus the model parameters for the \texttt{sp--s-i} set in this paper (that is, the model set that includes only disks around a central star, with no variable inner radius). Also shown for reference are the results for the \texttt{spubhmi} set (the most complex one presented in this paper).

The figure shows the significant difference between the original parameter space coverage, which is non-uniform and has a complex shape, and the new parameter space coverage, which is much more uniform in both model sets. The models that provide a good fit lie in a similar region of parameter space, but show more structure in the results from the R06 models, which could be due to biases in the parameter space coverage that may be in other dimensions that those shown here. In contrast, the new models that provide a good fit have a more uniform and extended distribution in the \texttt{sp--s-i} set, and show that the temperature and total disk mass are both very uncertain, although certain ranges are ruled out. The lack of strong constraints is expected -- without a detection in the sub-mm, the disk mass cannot be well constrained, and the mid-infrared fluxes can only provide a lower limit to the mass. Indeed, as shown in R06, mid-infrared fluxes are only able to tell us once the disk has become optically thick, but above this they provide no constraints. The stellar temperature is uncertain because of the large extinction -- to some extent, the temperature and extinction are in fact correlated.

\section{Caveats}

\label{sec:caveats}

While this new set of models addresses the limitations of the R06 models described in Section~\ref{sec:limitations}, there are nevertheless a number of caveats with the current models. Some of these were already mentioned above, and a summary is provided here. First, the caveats relating to parameter space coverage are the following:
\begin{enumerate}
\item A single dust model is assumed throughout the density structures (c.f. \S\ref{sec:dust}).
\item The dust properties do not take into account emission from transiently heated very small grains and polycyclic aromatic hydrocarbons (PAHs).
\item All the models have a single source of emission, which is the central star, and the effects of the interstellar radiation field are not taken into account.
\item Accretion is not explicitly included. As mentioned in Section~\ref{sec:disk}, this does not mean that the models cannot be used to study objects with accretion -- it mainly means that the models cannot reproduce the UV and optical fluxes for non-embedded objects.
\item Not all models are necessarily equally physically plausible. It is up to the user to decide whether certain models should be ignored for not being physically realistic.
\end{enumerate}
Future versions of the models can of course address these limitations, for example by adding models with different dust properties and even variable dust properties, but it is important to also have basic models that have simple dust properties and no interstellar radiation field for comparison, in order to test whether there is actually evidence that these simple properties are not sufficient. Other important caveats include:
\begin{enumerate}
\setcounter{enumi}{5}
\item The shape of the photosphere models for the central source may be accurate only to 10\% in some cases (as discussed in Section~\ref{sec:source}). This will only matter when modeling fluxes in the optical and near-infrared for non-embedded objects.
\item Parameters derived from the model parameters may still suffer from correlations, even though the main parameters are sampled uniformly inside ranges. For example, the distribution of radii and temperatures is uniform in log space, but the distribution of luminosities is not so. This is not an issue with the model sets presented here as such, but the limitation is that only the directly sampled model parameters are uniform (it would indeed be impossible to sample radii, temperatures, and luminosities evenly in the same model set).
\item The models should not be treated as final: similarly to software, the models can include issues that will be resolved in subsequent versions. For example, in the \texttt{v1.1} release of the models described in this paper, a small fraction (2\%) of the models are missing, and for two of the model sets, a small fraction of models will be missing scattered dust emission in the mid-infrared (see \S\ref{sec:computational} for more details). When publishing results based on these models, it is therefore important to mention the version used.
\end{enumerate}

\section{General limitations}

\label{sec:general_limitations}

In Section~\ref{sec:limitations}, I outlined limitations with the R06 set of models, which are addressed with the new models. However, there are a number of limitations of SED modeling in general that no set of models can overcome, and are instead intrinsic to this type of modeling.

\subsection{Misspecification of models}

\label{sec:misspecification}

YSOs, especially accreting protostars, are complex 3-d objects, not nicely axisymmetric density structures. Even disks may contain 3-d structures such as spiral arms. This means that from the onset, we know that the models are wrong compared to the real density distributions, but we can still hope that they are similar enough to reality to provide interesting insight into those objects.

In the language of statistics, our models are misspecified. Most models in Astrophysics are misspecified, but to a different degree: for instance, models for the orbits of binary stars, or for parallax motions, while still approximate, are much better approximations of reality than models of the density distribution around YSOs.

In practice, the fact the models are misspecified means that one cannot fit these models to data using $\chi^2$ minimization and expect that the $\chi^2$ value will be convertible into a probability using the classical $P\propto\exp{(-\chi^2/2)}$. Instead, the $\chi^2$ can only be used as a relative measure of goodness of fit -- the absolute value is not meaningful.

\subsection{Blending and confusion}

With the exception of the nearest and low-density star-forming regions, more crowded regions and more distant star-forming regions suffer from source blending, and many objects that appear as single YSOs will often in fact be two or more objects. If this is the case, then some of the bulk properties such as the luminosity or dust mass may be close enough to the real values (although the latter depends on the temperature distribution, which will be different if multiple sources are present), but for the more detailed parameters the values will not be meaningful.

For example, when modeling two blended sources, properties such as the disk flaring angle or scaleheight are no longer meaningful. Of course, it is not always possible to know if a source is suffering from blending, especially for distant star forming regions, but one should always bear in mind that this is likely, and therefore that the more distant regions one looks at, the less one should trust the details of the individual parameters, and for these regions, it may be more useful to carry out population synthesis modeling than to fit single SEDs to photometry which most likely contains contributions from several sources.

\subsection{Variability}

Even in the case of an isolated YSO, a further complication is (wavelength-dependent) variability. Many YSOs are known to be variable \citep[e.g.,][]{ysovar1, ysovar2}, and this would not be a problem in itself if all the multi-wavelength photometry was taken at the same time, but this is rarely the case. Most SEDs that we construct from archival data contain contributions at different wavelengths that were measured years apart in some cases. This can result in discontinuities in the SED from one wavelength to the next, making it challenging to fit the fluxes with a static model, but often not providing sufficient data to allow a detailed time-dependent model to be used.

\section{Summary}

\label{sec:summary}

I have presented a new set of model SEDs for young stellar objects, spanning a wide range of evolutionary stages, from the youngest deeply embedded protostars to pre-main-sequence stars with little or no disks. In Sect.~\ref{sec:limitations}, I discussed the limitations with the models that were originally published in R06, and in Sect.~\ref{sec:overview} I discussed how the new model set was designed in order to resolve these limitations. To summarize, the most significant changes compared to the previous generation of models are:

\begin{itemize}
\item The new models do not depend on highly model-dependent values, such as the stellar age and mass, which depend on stellar evolutionary tracks. Instead, the new models are defined using parameters that have a direct impact on the radiative transfer.
\item The parameter space is sampled uniformly (in linear or logarithmic space depending on the parameter) which does not introduce correlations between the parameters used to define the models.
\item The models are split into sets with increasing complexity, with the simplest model set having two parameters, and the most complex having 12.
\item The envelope outer radius for envelope models should now be large enough to include 10-20K dust that is essential for modeling \textit{Herschel} observations.
\item The models now have a high S/N over the entire wavelength range.
\end{itemize}

The aim of the new set of models is not to provide the most physically realistic models for young stars, but rather to provide deliberately simplified models for initial modeling, which allows us to cover a wider range of parameter space. In addition, the design of the new set of models allows us to separate the problem into determining constraints related to the presence or absence of components, and determining constraints related to the physical parameter values assuming a given model.

The fitting example in Section \ref{sec:fitting_examples} showed how these models can be used to analyze real observations of a YSO, and what we could determine from this. While the modeling results were highly degenerate, I showed how one can use a Bayesian approach to assign relative probabilities to the various models of increasing complexity.

The first version of the models (\texttt{v1.1}) as well as a new Python-based fitting tool are publicly available.

\section*{Acknowledgements}

\label{sec:acknowledgements}

I wish to thank Barbara Whitney and Kenny Wood for helpful discussions that helped shape this work, as well as the referee for suggestions that helped improve this paper. This research made use of Astropy, a community-developed core Python package for Astronomy \citep{astropy:13}. All the scripts used to produce the figures in this paper and to carry out the fitting example in Section \ref{sec:fitting_examples} will be made available alongside the models. Since I am no longer working as a researcher, the model sets presented here will not be expanded in future to include additional parameters (although I may release updated versions of the models to fix the issues described in \S\ref{sec:computational}), but all the materials needed to reproduce this work will be freely available so that anyone interested in expanding this work will be able to do so.

\bibliography{}

\end{document}